\documentclass[lettersize,journal]{IEEEtran}
\def\ps@headings{%
\def\@oddhead{\mbox{}\scriptsize\rightmark \hfil \thepage}%
\def\@evenhead{\scriptsize\thepage \hfil \leftmark\mbox{}}%
\def\@oddfoot{}%
\def\@evenfoot{}}
\makeatother 

\usepackage{tikz}
\usepackage{xurl}
\usepackage[outline]{contour}
\usepackage{enumitem}
\usepackage{url}
\usepackage{cite}
\usepackage{comment}
\usepackage{amsmath,amssymb,amsfonts}
\usepackage{algorithmic}
\usepackage{graphicx}
\usetikzlibrary{patterns}
\usepackage{textcomp}
\usepackage{balance}
\usepackage{xcolor}
\usepackage{bm}
\usepackage{array}
\usepackage{booktabs}
\usepackage{multirow}
\usepackage{subcaption}
\usepackage{bbm}
\usepackage[utf8]{inputenc}
\usepackage{array}
\usepackage{amsthm}
\usepackage{lipsum}

\usetikzlibrary{arrows.meta,calc,positioning}
\usepackage{balance}
\usepackage{float}
\usepackage{orcidlink}
\usepackage{chngpage}
\usepackage{hyperref}
\usepackage{pifont}
\usepackage[hyphenbreaks]{breakurl}
\usepackage[normalem]{ulem}

\def\beq{\begin{equation}}
\def\eeq{\end{equation}}
\def\bea{\begin{eqnarray}}
\def\eea{\end{eqnarray}}
\def\ba{\begin{array}}
\def\ea{\end{array}}

\def\bitem{\begin{itemize}}
\def\eitem{\end{itemize}}
\def\ben{\begin{enumerate}}
\def\een{\end{enumerate}}

\def\ie{{\it i.e.,\ \/}}

\definecolor{bgrd}{rgb}{1,1,1}
\definecolor{gray}{rgb}{0.5,0.5,0.5}
\definecolor{dkr}{rgb}{0.7,0.1,0.2}
\definecolor{dkb}{rgb}{0.1,0.1,0.8}

\def\tcb{\textcolor{blue}}

\def\H{\mbox{\tiny H}}

\newcommand{\mbbR}{\mathbb{R}}

\def\Bc{{\cal B}}

\def\Tc{{\cal T}}

\newtheorem{proposition}{Proposition}

\newtheorem{corollary}{Corollary}
\newtheorem{theorem}{Theorem}

\newcounter{subassumption}[assumption]

\definecolor{pvyellow}{RGB}{247,214,34}
\definecolor{chargegreen}{RGB}{150,230,170}
\definecolor{dischargegreen}{RGB}{42,220,255}
\definecolor{netgray}{RGB}{220,220,255}
\usepackage{pgfplots}
\usepackage{xcolor}
\usepackage[table]{xcolor}
\usepgfplotslibrary{fillbetween,groupplots}
\usetikzlibrary{arrows.meta}
\definecolor{pvblue}{RGB}{20,70,220}
\definecolor{storgreen}{RGB}{0,125,55}
\definecolor{cnorange}{RGB}{215,105,0}
\definecolor{pvyellow}{RGB}{247,214,34}
\definecolor{chargegreen}{RGB}{150,230,170}
\definecolor{dischargegreen}{RGB}{42,220,255}
\definecolor{netgray}{RGB}{220,220,255}
\definecolor{ArrUp}{HTML}{C0392B}
\definecolor{ArrDn}{HTML}{2874A6}
\definecolor{ArrNm}{HTML}{707B7C}
\definecolor{SecBg}{HTML}{ECF0F1}

\newcommand{\on}{\rm{on}}
\newcommand{\off}{\rm{off}}
\pgfplotsset{compat=1.18}
\usepackage{pgfplots}
\usepgfplotslibrary{groupplots}
\pgfplotsset{compat=1.18}

\makeatletter
\def\munderbar#1{\underline{\sbox\tw@{$#1$}\dp\tw@\z@\box\tw@}}

\DeclareRobustCommand{\tcb}[1]{\textcolor{blue}{#1}}
\def\BibTeX{{\rm B\kern-.05em{\sc i\kern-.025em b}\kern-.08em
    T\kern-.1667em\lower.7ex\hbox{E}\kern-.125emX}}

\title{When Does Storage Complement Solar? \\ Investment and Cost Recovery under Advanced Net Metering}

\author{Nathan Engelman Lado, Ahmed~S.~Alahmed\orcidlink{0000-0002-4715-4379},
Audun~Botterud\orcidlink{0000-0002-5706-5636},
Saurabh~Amin\orcidlink{0000-0003-1554-015X}%
\thanks{\scriptsize Nathan Engelman Lado, Audun Botterud, and Saurabh Amin ({\tt \{{\tcb{engelm25,audunb,amins}\}\tcb{@mit.edu}}}) are with the Massachusetts Institute of Technology, Cambridge, MA 02139 USA. Ahmed S. Alahmed ({\tt {\tcb{alahmad@kfupm.edu.sa}}}) is with the King Fahd University of Petroleum and Minerals, Dhahran 31261, Saudi Arabia. The full paper, including proofs of theoretical results, is available in \cite{lado&alahmed&botterud&amin:26JournalarXiv}. A preliminary version of this paper was published as a conference paper at the PowerUp Conference, Boulder, CO, 2026 \cite{lado&alahmed&botterud&amin:26PU}.}
}

\begin{document}
\begingroup
\allowdisplaybreaks

\maketitle

\begin{abstract}
We examine the joint PV and storage operational and investment decisions of a \textit{prosumer} under net energy metering (NEM) and peak/off-peak tariffs (ToU). Legacy NEM compensates net exports at the retail rate over a netting period, typically under time-invariant pricing. To better reflect system costs, tariff features are being reconsidered, with many U.S. states adopting time-varying prices and export compensation below the retail price. 
We formulate the prosumer decision problem as a cost minimization problem with charge/discharge, PV investment, and storage investment decisions. We analytically derive the optimal decisions based on the tariff’s membership in one of six regions that partition the tariff space. Using these solutions, we examine the relationship between PV and storage, identifying when they are \textit{complements} or \textit{substitutes}. Furthermore, we identify the relationship between private DER incentives and the distribution utility's (DU's) net cost of serving the prosumer. For both storage and PV, when investment costs are greater than the costs of serving the customer, there is no tariff where investment in PV and storage decrease non-adopter bills. Under the same conditions, complementarity between PV and storage decreases DU cost recovery for a given tariff. In a Massachusetts-based case study, we validate our theoretical findings by connecting tariff design features to DER interaction, the DU's net cost, and DER adoption in a stochastic, multi-prosumer, and multi-period setting.
\end{abstract}


\section{Introduction}
\IEEEPARstart{T}{he} rapid deployment of behind-the-meter (BTM) distributed energy resources (DERs), particularly rooftop photovoltaic (PV) systems, is transforming electricity consumption and generation patterns at the distribution level. Consumers equipped with such resources, termed prosumers, are able to both import electricity from and export surplus generation to the distribution utility (DU)~\cite{Burger&Perez&Arriaga:19CEEPR}. Net energy metering (NEM) has emerged as the predominant tariff mechanism for facilitating bidirectional energy exchange in the United States. Under conventional NEM, a prosumer's net energy consumption is calculated over a netting period. Net imports are charged at the retail price and net exports are compensated with credits, traditionally valued near the retail rate (symmetric NEM) or at a reduced price (asymmetric NEM)~\cite{NEMevolution:23NAS}%
\footnote{Symmetric and asymmetric NEM designs are also referred to in the literature as full NEM and partial NEM (or net billing), respectively.}.
{\color{black}
Legacy NEM tariffs are often implemented in conjunction with time-invariant (TI) volumetric tariffs. This means that residential prosumers are charged the same per \$/kWh rate no matter the time of day (time-invariance) or whether they are net-producers or consumers (symmetric NEM). Both TI pricing and symmetric NEM cause distortions to the price of electricity as they do not accurately convey the costs of supplying prosumers with electricity. TI pricing does not represent the actual cost of producing electricity in a particular time period. Symmetric NEM increases this distortion when export compensation exceeds the costs that the grid avoids from receiving that electricity.

A second problem of volumetric tariffs that is exacerbated by NEM is the recovery of fixed costs. 
Since fixed and per kW costs are being recovered through a per kWh charge, NEM not only lets the prosumer avoid the per kWh charge, 
it can also reduce their contribution to costs that do not decrease proportionally with their net purchases \cite{hledik:24LBNL}.

We focus on NEM tariffs because their import and export prices determine the private value of PV generation, and how storage changes these values.
Although many existing NEM tariffs remain TI and symmetric, a growing number of jurisdictions are transitioning toward time-varying and asymmetric import/export tariff structures. Notable examples of asymmetric NEM tariffs include those adopted in California, Arizona, Georgia, Michigan, and Illinois~\cite{DSIRE}. These reforms aim to align customer incentives with system costs~\cite{Next10Report2022,Alahmed&Tong:22ACMSEIR}. Lowering the export rate reduces over-compensation when export credits exceed avoided system costs. Similarly, shifting from a TI tariff to a time-of-use (ToU) tariff brings exposes the customers to a tariff that better reflects variation in system costs. Here, we examine a tariff with peak and off-peak periods, the simplest ToU schedule.

Asymmetric NEM and ToU can also encourage prosumers to pair storage with PV. When export compensation is below the retail price, storing surplus PV generation for later consumption can increase the private value of solar electricity. ToU prices also increase storage profitability since low-cost grid or stored solar generation displaces purchases during higher-priced periods. Due to this interaction, ToU tariffs are perceived as encouraging adoption of combined PV and storage systems~\cite{SepulvedaMora:21Energy} and peak reduction that provides both private and system benefits~\cite{Zakeri:21AE}. These potential benefits also motivate policy recommendations to use time-varying retail rates to compensate the system value of combinations of DERs~\cite{Shenot:19RAP}. However, the conditions under which PV and storage act as complements or substitutes under a NEM and whether it is possible to find an NEM tariff which reduces non-adopter bills while recovering costs have not been addressed (to the best of our knowledge).

 Much of the existing literature on NEM examines the problem through an econometric analysis of consumers and prosumers using historical data or by examining prosumer decision-making as an optimization problem, although there is overlap. In the first category, early work studies prosumer benefits, net demand, and PV adoption under legacy NEM tariffs with symmetric prices \cite{Borenstein:15NBER, Darghouth&Barbose&Wiser:11EP,Yamamoto:17EP}. Other work examines the effect of NEM on bill savings, cost shifts to non-adopting consumers, and the overall efficiency of NEM tariffs \cite{Burger&Perez&Arriaga:19CEEPR}. Another line of work within this category endogenizes the PV capacity under NEM, but abstracts from storage operation \cite{Comello&Reichelstein:17RSER,Cerino&Noussan:18JCP}. These works highlight the distributional and revenue-adequacy consequences of volumetric cost recovery when customers can adopt BTM generation but without considering the endogenous storage adoption.
 In the second category of literature, operational decisions and DER aggregation are presented as an optimization under time-varying and asymmetric NEM tariffs \cite{Alahmed&Tong:22TSG, Chakraborty&Poolla&Varaiya:19TSG}, but generally maintain the assumption of exogenous PV capacity. A separate line of work endogenizes the PV capacity under NEM, but abstracts from demand flexibility and storage operation \cite{Comello&Reichelstein:17RSER, Chakraborty&Khargonekar:23CSL, Cerino&Noussan:18JCP}.
Related literature studies how tariff design, DER adoption, and utility cost recovery interact at the aggregate level, while keeping DER adoption decisions fairly abstract \cite{Sun&Feng&Tong:20TAC}.

Our work is in the second category. We extend the literature on prosumer behavior and examine its implications for retail tariff design in order to address problems with NEM described by the first literature category. Specifically, we characterize how a prosumer jointly optimizes operational decisions, PV investment, and storage investment under a potentially asymmetric NEM tariff with peak and off-peak periods. We show when PV and storage are complements or substitutes under this tariff structure, and how these decisions impact the regulator tariff design problem.

\subsection*{Summary of Contributions:}
 This paper first studies the joint optimal investment and operational decisions of a prosumer with BTM PV generation and battery storage under ToU-NEM tariffs. In particular, the paper makes the following contributions:
\begin{itemize}[leftmargin=*, labelsep=0.5em]
\item \textbf{Optimal Decisions and Price Regions:}
We identify six regions that partition the ToU-NEM tariff space and show how optimal storage operation and investment is determined based upon tariff location within these regions. 

\item \textbf{Complementarity/Substitutability:}
We show that PV and storage act as complements or substitutes based on three factors: (1) whether storage causes a crossing of the net zero boundary, (2) the distribution of PV capacity factor, and (3) import/export price differences.

\item \textbf{DER interaction and cost recovery:} We provide sufficient conditions for when DERs increase DU costs. We then establish the two roles complementarity plays in tariff design: increasing DU costs and increasing PV investment. 
\end{itemize}
In summary, the main contribution of our work is to examine the interaction the joint investment of PV and storage operational connect this to the regulator's tariff design problem.

The paper proceeds as follows. Section \ref{sec:model} formulates the NEM tariff and the prosumer problem. Section \ref{sec:optimal_decisions} characterizes battery operations, PV investment, and storage investment decisions under the tariff. Section \ref{sec:system} analyzes the DU net-cost effect of PV and storage adoption. Section \ref{sec:CaseStudy} presents a Massachusetts-based case study, and Section \ref{sec:conclusion} concludes. Formal proofs are provided in the appendix in \cite{lado&alahmed&botterud&amin:26JournalarXiv}.

\section{Prosumer Model and Decision Problem} \label{sec:model}
\subsection{Prosumer DERs and Payment Under ToU-NEM}
We consider a prosumer with  rooftop PV generation and battery energy storage capabilities, receiving service from a DU under a ToU-NEM tariff structure. 
The prosumer jointly selects DER investment levels and operating decisions over a finite set of operational, periods $t  \in  \Tc$ under perfect foresight. The operating periods alternate between off-peak periods, $\Tc_{\off}$ and peak periods $\Tc_{\on}$, with $T=|\Tc_{\on}|=|\Tc_{\off}|$. Each off-peak period is followed by a peak period, so $T$ is also the number of charge-discharge cycles available to storage. Although physical energy imports and exports occur continuously, the prosumer's settlement with the DU is based upon their net demand over the peak and off-peak netting periods $b \in \Bc:=\{\off,\on\}$ with $d_b^+, d_b^-  \in  \mathbb{R}_+$ denoting \textit{energy import} and \textit{export}, respectively. Without loss of generality, we assume that the first operational period is a member of the off-peak netting period.

For every period $t\in \Tc$, we denote the prosumer’s {\em electricity consumption} by $d_{t} \in  \mathbb{R}_+$, {\em PV electricity generation} by $\psi_t g$, and {\em battery charging}, {\em discharging}, and {\em state-of-charge} (SoC) by $q_t^c \in  \mathbb{R}_+$, $q_t^d \in  \mathbb{R}_+$, and $s_t \in  \mathbb{R}_+$, respectively. PV generation is the product of the capacity factor $\psi_t \in  \mbbR_+$ and the PV capacity $g \in [0,\bar g]$, with $\bar g>0$. We denote the total capacity factor over netting period $b\in\{\on,\off\}$ as $\Psi_b =\sum_{t\in\mathcal{T}_b}\psi_t$. Storage has energy capacity $z \in [0,\bar z]$ and charging and discharging efficiencies $\eta^c,\eta^d \in (0,1)$ with $\bar g,\bar z>0$. We do not include battery power capacity as we assume that the peak and off-peak periods are long enough that power capacity does not constrain battery operation. The annualized marginal investment costs of a kW of PV and a kWh of storage capacity are  $c^g\!\in\! \mathbb{R}_+$ and $c^z\!\in\! \mathbb{R}_+$, respectively.  Consumption is exogenous and price-inelastic, while capacity factors are deterministic and known by the prosumer. In Section \ref{sec:optimal_decisions}, these modeling choices allow us to characterize optimal operation and investment in closed form, providing significant insight into the relationships between DERs and tariffs. In our case study (Sec. \ref{sec:CaseStudy}), we extend our model by using stochastic capacity factors and a greater number of netting in our case study.



\subsection{The ToU-NEM Tariff}
We study a prosumer under a ToU-NEM tariff that permits distinct prices across peak/off-peak periods and between net energy import and export, generalizing both symmetric NEM (which equates import and export prices) and TI-NEM pricing (which equates prices across the peak and off-peak periods). 
 Under ToU-NEM, both import and export prices can vary across periods to reflect differences in system net demand and the availability of low-cost generation.

Using the import and export variables defined for the peak and off-peak netting periods, the total payment under NEM, $P^{\mbox{\tiny NEM}} \in \mbbR$, is characterized by the tariff $\bm\pi:= (\{\pi_b^+\}_{b\in\mathcal B}, \{\pi_b^-\}_{b\in\mathcal B}, \pi^{\rm{fix}})$ and the per-period imports and exports $\bm{d}^+\!:=\!\{d_b^+\}_{b\in\mathcal{B}}$, $\bm{d}^-\!:=\!\{d_b^-\}_{b\in\mathcal{B}}$
as
\begin{align}
P^{\mbox{\tiny NEM}}(\bm d^+,\bm d^-; \bm \pi)
&:=
\sum_{b\in\mathcal B}(\pi_b^+\,d_{b}^+ - \pi_b^-\,d_{b}^-) + \pi^{\rm{fix}},\label{eq:Pi_NEM_def}
\end{align}
\vspace{-6pt}

\noindent where $\pi_b^+\in \mathbb{R}_+, \pi_b^- \in \mathbb{R}_+$ are the import (retail) and export (credit) prices, and $\pi^{\rm{fix}} \in \mathbb{R} $ is the fixed (connection) charge.\footnote{Such charges typically bundle non-fixed components such as network capacity costs.} We assume that the import price is weakly greater than the export price, $\pi_b^+\geq\pi_b^-$. This price ordering holds in practice, except in places that adopt buy-all-sell-all tariffs, which prohibit self-consumption of local generation \cite{Alahmed&Tong:22ACMSEIR}.

We define \textit{prosumer-specific} peak period as the netting period with higher prosumer gross demand relative to PV capacity factor, $\sum_{t\in\mathcal T_b}d_t/\Psi_b$. We consider the case where prosumer and system peaks coincide and a smaller fraction of prosumer demand can be offset by PV in the peak relative to the off-peak, \ie $\sum_{t\in\mathcal T_{\on}}d_t/(\Psi_{\on})>\sum_{t\in\mathcal T_{\off}}d_t/(\Psi_{\off})$, the corresponding off-peak prices no less than the corresponding off-peak prices: $\pi^{+}_{\on}\geq \pi^{+}_{\off}, \pi^{-}_{\on}\geq \pi^{-}_{\off}$. Therefore, the notation $\on,\off$ refers to both system and prosumer-specific peak and off-peak periods. We also define net demand without storage as $D_b(g):=\sum_{t\in\mathcal{T}_b}(d_t-\psi_t g)$. We assume PV can reduce peak imports but not eliminate them unless storage is installed, $D_{\on}(\bar g)> 0$. Together PV and storage can eliminate net imports in both periods, \ie, there exists a $z\in[0,\bar z]$ such that $D_{\off}(\bar g)+Tz/\eta^c\leq 0$ and $D_{\on}(\bar g)-\eta^dTz\leq 0$. Finally, we assume that, at maximum PV capacity, total off-peak surplus generation is sufficient to fully charge the largest admissible battery $T$ times, \ie $D_{\off}(\bar g)+T\bar z/\eta^c\leq0$.

\subsection{Prosumer Decision Problem}
We model the prosumer’s decisions using a cost minimization problem, where cost is defined as the charges associated with consumption and cost of investment decisions minus revenue derived from exported energy.
The prosumer minimizes its cost, $\Pi(g,z;\bm\pi)$, by optimally sizing its DERs based on optimal operational charge, discharge, import, and export decisions:
\begin{subequations}\label{eqn:prosumer_prob}
\begin{align}
&\min_{g,z}\Pi(g,z;\bm\pi)\nonumber
\\&=\min_{g,z}\big\{
c^z z+c^g g
+\min_{\substack{
d_b^+,d_b^-,\\
q_t^c,q_t^d,s_t}}
P^{\mbox{\tiny NEM}}(\bm d^+,\bm d^-;\bm\pi)
\big\}
\label{eqn:prosumer_opt}
\end{align}
\begin{align}
\text{s.t.}\,
& \sum_{t\in\mathcal T_b}
\!(
d_{t}\!-\! \psi_{t} g\!-\! q^d_{t}\!+\! q^c_{t})
\!=\! d_{b}^+ \!-\! d_{b}^-
&&\!\! \forall b \in \mathcal{B}
\label{eqn:cons_balance}
\\[-6pt]
& s_{t+1}
= s_{t}
+ \eta^c q^c_{t}
- \frac{1}{\eta^d} q^d_{t}
&&\!\!\forall t \in \mathcal{T}
\label{eqn:soc_dyn}
\\[-4pt]
& 0 \le s_{t} \le z 
&&\!\! \forall t \in \mathcal{T}
\label{eqn:soc_bounds}\\
&s_{T+1}=s_1,
\label{eqn:soc_cycle}
\\
&  0 \le z\le \bar z \label{cons:stor_cap}
\\
& 0 \le g \le \bar g \label{cons:pv_cap}
\\
& d_{b}^+ \ge 0
\quad d_{b}^- \ge 0
&&\!\!\forall b \in \mathcal{B}\label{cons:impexp_noneg}\\
& q^c_{t} \ge 0,\quad q^d_{t} \ge 0
 &&\!\!\forall t \in \mathcal{T} 
\label{cons:chg_noneg}\\
& d_{b}^+ \cdot d_{b}^- = 0
&&\!\! \forall b \in \mathcal{B}.
\label{cons:bilinear}\\
& q_{t}^d \cdot q_{t}^c = 0
&&\!\! \forall t \in \mathcal{T}.
\label{cons:bilinear_str}
\end{align}\end{subequations}
The objective in \eqref{eqn:prosumer_opt} consists of NEM payments as well as costs associated with PV investment $g$  and storage investment $z$.
 The balance constraint \eqref{eqn:cons_balance} links net consumption within each netting period to net grid transactions. The SoC dynamics in \eqref{eqn:soc_dyn} track the quantity of stored energy, accounting for charging and discharging, efficiencies. Constraint \eqref{eqn:soc_cycle} ensures that the battery to begin and end with the same SoC, while \eqref{eqn:soc_bounds} limits the SoC by installed storage capacity. Constraints \eqref{cons:pv_cap} and \eqref{cons:stor_cap} bound storage and PV investments with the maximum possible investments. Constraints \eqref{cons:impexp_noneg} and \eqref{cons:chg_noneg} impose non-negativity on operational decisions.

The bilinear constraints in \eqref{cons:bilinear}-\eqref{cons:bilinear_str} prohibits simultaneous importing and exporting as well as simultaneous charge and discharging. Though this introduces non-convexity, when the import price is no less than export prices, $\pi^+_b\!\geq\!\pi^-_b \forall t$ and $\eta^c\eta^d<1$, there exists an optimal solution such that \eqref{cons:bilinear} and \eqref{cons:bilinear_str} are satisfied.
This is formalized in Lemma S.1 (See the appendix in \cite{lado&alahmed&botterud&amin:26JournalarXiv}), which we use to solve \eqref{eqn:prosumer_opt}-\eqref{cons:chg_noneg} without the bilinear constraints. The remaining problem is a linear optimization program. We examine this linear program in the following section to derive the prosumer's optimal operational and investment decisions under NEM.

\section{Optimal DER Operational and Investment Decisions} \label{sec:optimal_decisions}
This section characterizes the prosumer’s optimal operational and investment decisions under a ToU-NEM tariff, and the conditions under which PV and storage are complements or substitutes.

\subsection{Optimal Operational Decisions}
We begin with optimal operational decisions, holding $g$ and $z$ fixed. We need only solve for optimal charge and discharge as net demand and SoC can be expressed as functions of charging decisions through \eqref{eqn:cons_balance} and \eqref{eqn:soc_dyn}, respectively. 

For each $b \in \mathcal B$, $\gamma_b^*$ denotes the subderivative of $\Pi(g,z;\bm \pi)$ with respect to the charge. This subderivative is determined by the operating regime of the netting period $b$: 
\begin{align} \gamma_b^* \in \begin{cases} \{\pi_b^+\}, & d_b^{+*}>0,\hspace{40pt}\textit{Import}\\ \{\pi_b^-\}, & d_b^{-*}>0,\hspace{40pt}\textit{Export}\\ [\pi_b^-,\pi_b^+], & d_b^{+*}+d_b^{-*}=0. \hspace{13pt}\textit{net zero}\end{cases} \label{eq:gamma_regime_characterization} \end{align}
The marginal value of storing one unit of energy off-peak and releasing it in the following peak period is $\eta^c\eta^d\gamma_{\on}-\gamma_{\off}$. Thus, battery operation depends on whether off-peak energy comes from surplus generation or imports and whether peak discharge offsets imports or creates exports. The threshold separating these regimes occurs when total off-peak charge is $-D_{\off}(g)$ and $D_{\on}(g)/(\eta^c\eta^d)$. Under $\pi_b^+\geq\pi_b^-$, these tariff component comparisons yield the six price regions, which we label A-F, used in the following proposition:
\begin{proposition}[Optimal operational decisions]
\label{prop:conditional_operating_billing}

Fix PV and storage capacities $g$ and $z$. Then for $t\in\mathcal{T}_{\off}$, the charging decision $q_t^{c*}=\max\{\min\{Q_{\off}^{c\dagger}/T,z/\eta^c\},0\}$ is optimal, where 
\begin{align} 
&Q^{c\dagger}_{\off}(g;\bm\pi)\label{eq:qdagger_price_regimes_fixed_demand}
\\&=\!
\begin{cases}
\!0,
&
\!\!\!\!\!\!\!\pi_{{\on}}^+\!<\! \frac{\pi_{{\off}}^-}{\eta^c\eta^d} \hspace{55pt} \textbf{(\rm A)},
\\
\!\min\{
-D_{{\off}}(g),\!
\dfrac{D_{{\on}}(g)}{\eta^c\eta^d}\!
\},
&
\!\!\!\!\!\!\pi_{{\on}}^-
\!<\!
\frac{\pi_{{\off}}^-}{\eta^c\eta^d} \!\leq\!
\pi_{{\on}}^+ \!<\!
\frac{\pi_{{\off}}^+}{\eta^c\eta^d}  \textbf{(\rm B)},
\\
\!\dfrac{D_{{\on}}(g)}{\eta^c\eta^d},
&
\!\!\!\!\!\!\pi_{{\on}}^- \!<\!
\frac{\pi_{{\off}}^-}{\eta^c\eta^d} \!\leq\!
\frac{\pi_{{\off}}^+}{\eta^c\eta^d} \!\leq\!
\pi_{{\on}}^+ \textbf{(\rm C)},
\\
-\!D_{{\off}}(g),
&
\!\!\!\!\!\!\!\frac{\pi_{{\off}}^-}{\eta^c\eta^d}
\!\le\!
\pi_{{\on}}^-
\!\leq\!
\pi_{{\on}}^+
\!\leq\!
\frac{\pi_{\off}^+}{\eta^c\eta^d} \hspace{1pt}\textbf{(\rm D)},
\\
\!\max\{\!
-\!D_{{\off}}(g),
\!\dfrac{D_{{\on}}(g)}{\eta^c\eta^d}
\!\} ,
&
\!\!\!\!\!\!\!\frac{\pi_{{\off}}^-}{\eta^c\eta^d}
\!\le\!
\pi_{{\on}}^-
\!<\!
\frac{\pi_{{\off}}^+}{\eta^c\eta^d}
\!\leq\!
\pi_{{\on}}^+ \hspace{1pt}\textbf{(\rm E)},
\\
\!+\infty,
&
\!\!\!\!\!\!\!\frac{\pi_{{\off}}^+}{\eta^c\eta^d}\le
\pi_{{\on}}^-.\hspace{49pt}\textbf{(\rm{F})}
\end{cases}\nonumber
\end{align}
The corresponding optimal battery decisions are $q_{t'}^{d*}\!\!=\!q_t^{c*}\eta^c\eta^d$ and $q_{t'}^{c*}=q_t^{d*}=0$ for every $t'  \in \mathcal{T}_{\on}$ and $t\in \mathcal{T}_{\off}$ and net demand is determined by \eqref{eqn:cons_balance}. Proof is provided in \cite{lado&alahmed&botterud&amin:26JournalarXiv}.
\end{proposition}
\noindent Prop.~\ref{prop:conditional_operating_billing} characterizes off-peak charge implied by the tariff ordering and the corresponding capacity constraints. Then net demand, imports, and exports are determined by substituting battery operations into \eqref{eqn:cons_balance}. The operating regions follow from the principle that the battery charges until the efficiency-adjusted marginal value of peak discharge no longer exceeds the marginal opportunity cost of off-peak energy. Fig.~\ref{fig:charge_regimes_tariff_plane} displays the regions shown in Prop.~\ref{prop:conditional_operating_billing} as well as their respective capacity unconstrained optimal charging decisions. Peak import and export prices are varied so that their value relative to the off-peak prices defines the regions.
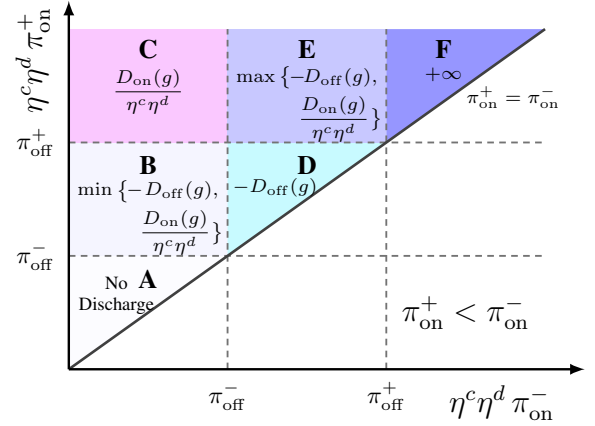
\begin{figure}[t]
\centering

\makebox[\columnwidth][c]{%
\scalebox{1}{%
\begin{tikzpicture}[
  x=2.1cm, y=1.5cm, >=latex, font=\small,
  grid/.style={dash pattern=on 2.5pt off 2pt, line width=0.7pt, black!55},
  axis/.style={->, line width=0.9pt},
  diag/.style={line width=1.0pt, black!75},
  infeas/.style={pattern=north east lines, pattern color=black!16},
  reglab/.style={font=\bfseries},
  rule/.style={font=\scriptsize},
  axlab/.style={font=\bfseries\normalsize}
]

\definecolor{regA}{RGB}{252,252,255}
\definecolor{regB}{RGB}{245,245,255}
\definecolor{regC}{RGB}{250,200,255}
\definecolor{regD}{RGB}{200,250,255}
\definecolor{regE}{RGB}{200,200,255}
\definecolor{regF}{RGB}{150,150,255}

\def\L{1}
\def\H{2}
\def\M{3}

\fill[white] (\L,0) rectangle (\M,\L);
\fill[white] (\H,\L) rectangle (\M,\H);

\fill[regA] (0,0) -- (\L,\L) -- (0,\L) -- cycle;
\fill[regB] (0,\L) rectangle (\L,\H);
\fill[regC] (0,\H) rectangle (\L,\M);
\fill[regD] (\L,\L) -- (\H,\H) -- (\L,\H) -- cycle;
\fill[regE] (\L,\H) rectangle (\H,\M);
\fill[regF] (\H,\H) -- (\M,\M) -- (\H,\M) -- cycle;

\draw[grid] (\L,0) -- (\L,\M);
\draw[grid] (\H,0) -- (\H,\M);
\draw[grid] (0,\L) -- (\M,\L);
\draw[grid] (0,\H) -- (\M,\H);
\draw[diag] (0,0) -- (\M,\M);

\draw[axis] (0,0) -- (\M+0.25,0);
\draw[axis] (0,0) -- (0,\M+0.25);

\node[axlab, anchor=north] at (\M/1.1,-0.06)
    {\large $\eta^c\eta^d\,\pi^-_{\on}$};

\node[axlab, rotate=90, anchor=south] at (-0.06,\M/1.1)
    {\large$\eta^c\eta^d\,\pi^+_{\on}$};

\node[anchor=north] at (\L,-0.03)
    {$\pi^-_{\off}$};

\node[anchor=north] at (\H,-0.03)
    {$\pi^+_{\off}$};

\node[anchor=east] at (-0.03,\L)
    {$\pi^-_{\off}$};

\node[anchor=east] at (-0.03,\H)
    {$\pi^+_{\off}$};

\node[reglab] at (0.50,0.8) {A};
\node[rule, align=center] at (0.30,0.65) { No \\ Discharge};

\node[reglab] at (0.50,1.80) {B};
\node[rule, align=center] at (0.50,1.34)
    {$\min\big\{\!\!-\!D_{\off}(g),$\\[3pt]
     \hspace{25pt}$\dfrac{D_{\on}(g)}{\eta^c\eta^d}\big\}$};

\node[reglab] at (0.50,2.82) {C};
\node[rule] at (0.50,2.45)
    {$\dfrac{D_{\on}(g)}{\eta^c\eta^d}$};

\node[reglab] at (1.5,1.80) {D};
\node[rule] at (1.3,1.6)
    {$-D_{\off}(g)$};

\node[reglab] at (1.50,2.82) {E};
\node[rule, align=center] at (1.50,2.34)
    {$\max\big\{\!\!-\!\!D_{\off}(g),$\\[3pt]
     \hspace{25pt}$\dfrac{D_{\on}(g)}{\eta^c\eta^d}\big\}$};

\node[reglab] at (2.36,2.82) {F};
\node[rule] at (2.36,2.60)
    {$+\infty$};

\node[rule, black, align=center] at (2.5,.5)
    {\large $\pi^+_{\on}<\pi^-_{\on}$};

\node[rule, black, align=center] at (2.8,2.4)
    { $\pi^+_{\on}=\pi^-_{\on}$};

\end{tikzpicture}
}
}

\caption{\small
Unconstrained off-peak charge $Q^{c\dagger}_{\off}$.
Moving left to right, peak \emph{exports} increase in value relative to off-peak imports and exports. Moving bottom to top, peak \emph{imports} increase in value relative to off-peak imports and exports. The white region violates $\pi^+_{\on}<\pi^-_{\on}$.
}
\label{fig:charge_regimes_tariff_plane}
\end{figure}

The regions in Fig.~\ref{fig:charge_regimes_tariff_plane} differ according to which sources of off-peak energy and uses of peak discharge are profitable. In Region A, using off-peak surplus that would otherwise be exported to displace peak imports is unprofitable, so the battery does not charge. In Region B, this trade is profitable, but charging stops when either the off-peak surplus, $-D_{\off}(g)$, or the peak deficit, $D_{\on}(g)/(\eta^c\eta^d)$, is exhausted. In Region C, grid energy imported off-peak can also profitably displace peak imports, so charging continues until the peak deficit is eliminated. In Region D, stored energy can profitably create peak exports, while additional off-peak imports are unprofitable, so charging continues until the off-peak surplus is exhausted.

In Region E, both extensions are profitable, so charging continues until both limits are exhausted, yielding $\max\{-D_{\off}(g),D_{\on}(g)/(\eta^c\eta^d)\}$. Beyond that point, the marginal unit would be purchased off-peak solely to create an additional peak export. In Region F, even grid energy imported off-peak can be profitably exported during the peak period, so the unconstrained charging target is unbounded and actual operation is limited only by battery capacity.

The results for optimal investment in the next subsection use the same map $Q^{c\dagger}_{\off}$ at storage-cost-adjusted prices. The six regions used for operational decisions, shifted to account for investment costs, also organize optimal investment.

\subsection{Optimal Investment Decisions}
\label{sec:invest}
The interval in \eqref{eq:gamma_regime_characterization} reflects that an increase in net demand leads to positive imports, while a decrease leads to positive exports. Using $\gamma^*_b$, we define the marginal value of PV capacity, and the directional marginal values as,
\begin{align}
\Phi(g;z,\bm\pi)
&\!=\!\sum_{b\in\mathcal B}\gamma_b^*\Psi_b,
\label{eq:pv_marginal_value_bounds}\\
\underline \Phi(g;z,\bm\pi)
&\!:=\!
\min_{\bm\gamma^*}\Phi(g;z,\bm\pi),\ 
\overline \Phi(g;z,\bm\pi)\!:=\!\max_{\bm\gamma^*}\Phi(g;z,\bm\pi).
\nonumber
\end{align}
Where $\Pi(g,z;\bm\pi)$ is differentiable, $\Phi(g;z,\bm\pi)=\underline \Phi(g;z,\bm\pi)=\overline \Phi(g;z,\bm\pi)$. For fixed storage capacity,  $\min_{g,z}\Pi(g,z;\bm\pi)$ is convex in PV capacity, so both one-sided marginal values of PV are weakly decreasing in $g$.

In \eqref{eq:pv_marginal_value_bounds}, the marginal value of PV depends on the sum of $\gamma_b^*$ across netting periods, weighted by PV production. 

Optimal interior PV investment satisfies
\begin{subequations}
\begin{align}
c^g&\in[\underline{\Phi}(g^*;z^*,\bm\pi),\overline{\Phi}(g^*;z^*,\bm\pi)],\label{eq:pv_opt}
\end{align}
\end{subequations}
so that the marginal PV cost lies between the one-sided marginal private values. To find the optimal PV and storage values, we calculate the marginal value of PV for each pair of off-peak and peak net demand regimes, indicated by the first and second superscript entries, respectively. The labels $+$, $-$, and $0$ denote \textit{import}, \textit{export}, and \textit{net zero}. For $\alpha,\beta\in\{-,+\}$, we define
\begin{subequations}\begin{align}
\Phi^{\alpha\beta}&\!:=
\pi_{\off}^{\alpha}\Psi_{\off}+
\pi_{\on}^{\beta}\Psi_{\on},\label{eqn:alpha_beta_MV}
\\
\Phi^{0\beta}&\!:=\pi_{\on}^{\beta}\Big(\eta^c\eta^d
\Psi_{\off}+
\Psi_{\on}\Big)-\frac{c^z\eta^c}{T}
\Psi_{\off},\label{eqn:zero_beta_MV}
\\
\Phi^{\alpha0}
&\!:=
\pi_{\off}^{\alpha}\Big(\Psi_{\off}\!
+\frac{1}{\eta^c\eta^d}\Psi_{\on}\!\Big)
+\frac{c^z}{\eta^dT}\Psi_{\on}.\label{eqn:alpha_zero_MV}
\end{align}\end{subequations}
 The value $\Phi^{\alpha\beta}$ applies when neither period is at net zero, while $\Phi^{0\beta}$ and $\Phi^{\alpha 0}$ account for the storage adjustment required to keep the indicated period at net zero. After optimizing operation and storage investment, the resulting cost is convex in $g$, so PV's marginal private value is weakly decreasing. The marginal values change at the following PV capacities:
\begin{subequations}\begin{align}g^{\off}\!(z)\!&=\! \frac{D_{\off}(0)\!+\!Tz/\eta^c}{\Psi_{\off}},\hspace{9pt}
 g^{\on}\!(z)\!=\!\frac{D_{\on}(0)\!-\!\eta^dTz}{\Psi_{\on}},\label{eqn:glevel_1}\\ g^{\rm{nz}}\! &=\!\frac{D_{\on}(0)+\eta^c\eta^dD_{\off}(0)}{\Psi_{\on}\!+\!\eta^c\eta^d\Psi_{\off}},\label{eqn:glevel_2}\end{align}\end{subequations}
where $g^{\off}(z)$ and $g^{\on}(z)$ are the PV capacities at which off-peak or peak net demand equals zero given full charge and discharge of storage capacity $z$. Greater storage increases $g^{\off}(z)$ and lowers $g^{\on}(z)$. The capacity $g^{\rm{nz}}$ is the PV level at which storage can bring both periods to net zero.

The private value generated by charging in the off-peak and discharging in the peak must be large enough to recover the cost of storage investment. To account for this, we define the storage cost adjusted price 
$\widehat{\pi}_{\on}^\pm =\pi_{\on}^\pm-c^z/(\eta^dT)$ (we could have adjusted the off-peak tariff components instead). Using this adjusted price, the net demand regime-specific marginal value \eqref{eqn:alpha_beta_MV}-\eqref{eqn:alpha_zero_MV}, and the PV investment levels in \eqref{eqn:glevel_1}-\eqref{eqn:glevel_2}, we express optimal PV capacity by price region and by comparing the investment cost to regime-specific marginal values:

\begin{proposition}[Jointly optimal PV and storage investment]
\label{prop:joint_pv_storage_investment}
Suppose that the adjusted prices lie in the strict interior of one of Regions A-F. If $c^g$ differs from every $\Phi$ value in that region, the optimal PV capacity $g^*(\bm\pi,\bar g,\bar z)$ is given by Table \ref{tab:optimal_PV}, with each displayed PV capacity clipped to $[0,\bar g]$.
In every region, the jointly optimal storage capacity is
\begin{align}
&z^*(\bm\pi,\bar g, \bar z)\label{eq:joint_storage_from_qdagger}\\&=\!\max\{\min\{\eta^c Q_{\off}^{c\dagger}(g^*\!;\!\{\widehat{\pi}_{\on}^+,\widehat{\pi}_{\on}^-,\pi_{\off}^+,\pi_{\off}^-\})/T,\bar z\},\!0\}.
\nonumber
\end{align}
Whenever $c^g$ equals a $\Phi$ value in that region, every PV capacity between the corresponding adjacent clipped entries is jointly optimal with the storage capacity given by \eqref{eq:joint_storage_from_qdagger}. Proof is provided in \cite{lado&alahmed&botterud&amin:26JournalarXiv}.
\end{proposition}

\begin{table}[t]
\centering
\setlength{\tabcolsep}{1.0pt}
\resizebox{\columnwidth}{!}{%
\begin{tabular}{|@{}c|c|c||c|c|c@{}|}
\hline
Reg.
& Cond. on $c^g$
& $g^*$
& Reg.
& Cond. on $c^g$
& $g^*$
\\ \hline\hline
\textbf{(A)}
&$c^g>\Phi^{\scriptscriptstyle ++}$
&$0$
&\textbf{(D)}
&$c^g>\Phi^{\scriptscriptstyle ++}$
&$0$
\\
&$\Phi^{\scriptscriptstyle ++}>c^g>\Phi^{\scriptscriptstyle -+}$
&$g^{\off}(0)$
&
&$\Phi^{\scriptscriptstyle ++}>c^g>\Phi^{\scriptscriptstyle 0+}$
&$g^{\off}(0)$
\\
&$c^g<\Phi^{\scriptscriptstyle -+}$
&$\bar g$
&
&$\Phi^{\scriptscriptstyle 0+}>c^g>\Phi^{\scriptscriptstyle 0-}$
&$g^{\rm{nz}}$
\\
&&&
&$\Phi^{\scriptscriptstyle 0-}>c^g>\Phi^{\scriptscriptstyle --}$
&$g^{\off}(\bar z)$
\\
&&&
&$c^g<\Phi^{\scriptscriptstyle --}$
&$\bar g$
\\ \hline
\textbf{(B)}
&$c^g>\Phi^{\scriptscriptstyle ++}$
&$0$
&\textbf{(E)}
&$c^g>\Phi^{\scriptscriptstyle ++}$
&$0$
\\
&$\Phi^{\scriptscriptstyle ++}>c^g>\Phi^{\scriptscriptstyle 0+}$
&$g^{\off}(0)$
&
&$\Phi^{\scriptscriptstyle ++}>c^g>\Phi^{\scriptscriptstyle +0}$
&$[g^{\on}(\bar z)]^+$
\\
&$\Phi^{\scriptscriptstyle 0+}>c^g>\Phi^{\scriptscriptstyle -0}$
&$g^{\rm{nz}}$
&
&$\Phi^{\scriptscriptstyle +0}>c^g>\Phi^{\scriptscriptstyle 0-}$
&$g^{\rm{nz}}$
\\
&$c^g<\Phi^{\scriptscriptstyle -0}$
&$\bar g$
&
&$\Phi^{\scriptscriptstyle 0-}>c^g>\Phi^{\scriptscriptstyle --}$
&$g^{\off}(\bar z)$
\\
&&&
&$c^g<\Phi^{\scriptscriptstyle --}$
&$\bar g$
\\ \hline
\textbf{(C)}
&$c^g>\Phi^{\scriptscriptstyle ++}$
&$0$
&\textbf{(F)}
&$c^g>\Phi^{\scriptscriptstyle ++}$
&$0$
\\
&$\Phi^{\scriptscriptstyle ++}>c^g>\Phi^{\scriptscriptstyle +0}$
&$[g^{\on}(\bar z)]^+$
&
&$\Phi^{\scriptscriptstyle ++}>c^g>\Phi^{\scriptscriptstyle +-}$
&$[g^{\on}(\bar z)]^+$
\\
&$\Phi^{\scriptscriptstyle +0}>c^g>\Phi^{\scriptscriptstyle -0}$
&$g^{\rm{nz}}$
&
&$\Phi^{\scriptscriptstyle +-}>c^g>\Phi^{\scriptscriptstyle --}$
&$g^{\off}(\bar z)$
\\
&$c^g<\Phi^{\scriptscriptstyle -0}$
&$\bar g$
&
&$c^g<\Phi^{\scriptscriptstyle --}$
&$\bar g$\\
\hline
\end{tabular}
}
\caption{\small Optimal PV capacities by price region and PV investment cost. The notation $[g^{\on}(\bar z)]^+$ sets a negative threshold to zero. Within each region, when $c^g$ equals the marginal value separating two cases, both listed PV capacities and every capacity between them are optimal.}
 \label{tab:optimal_PV}
\end{table}
\noindent  Prop. \ref{prop:joint_pv_storage_investment} expresses the investment problem as a sequence of boundary comparisons. Within each two-period net demand regime, PV's marginal private value is constant and given by the applicable $\Phi$ value. Hence, $g^*$ is the boundary at which $c^g$ lies between the one-sided marginal values, while $z^*$ adjusts according to \eqref{eq:joint_storage_from_qdagger}. As $c^g$ falls, the optimum moves weakly through the boundaries in Table~\ref{tab:optimal_PV}. 
For a given PV capacity, optimal storage capacity is less than optimal capacity unconstrained charge entering the battery, as capacity incurs the additional cost of investment. Therefore, the optimal storage capacity is fully utilized, so that $Q_{\off}^{c*}=Tz^*/\eta^c$.
\noindent

\noindent\textbf{\textit{PV-Storage Interaction}:} From Prop. \ref{prop:joint_pv_storage_investment} it is not immediately clear when PV and storage are complements and when they are substitutes. Charging can raise PV's off-peak marginal value, while discharging can lower its peak marginal value; their net effect determines whether PV and storage are complements, substitutes, or neutral.

Figure~\ref{fig:comp_sub} illustrates cases of substitution and complementarity. On the left-hand side, storage raises off-peak net demand and shifts the prosumer from exporting to importing. This increases the marginal value of PV by
$(\pi_{{\off}}^+-\pi_{{\off}}^-)\Psi_{\off}$ as now the additional unit would offset imports rather than exports.
In the right-hand panel, additional discharge moves the prosumer from importing to exporting, lowering $\gamma_{\on}$ from $\pi_{{\on}}^+$ to $\pi_{{\on}}^-$. This decreases the marginal value of PV by $ (\pi_{{\on}}^+-\pi_{{\on}}^-) \Psi_{\on} $, as it lowers the value of additional generation in the peak period.
\begin{figure}[t]
    \centering
  
\begin{tikzpicture}[
    x=0.7cm,y=1cm,
    axis/.style={black,line width=0.8pt},
    split/.style={black,line width=0.8pt},
    cons/.style={black,line width=1.4pt},
    netline/.style={black!30,line width=1.6pt},
    pvline/.style={pvyellow,line width=2pt},
    ndem/.style={pattern=north east lines, pattern color=black!70},
    export/.style={pattern=north west lines, pattern color=black!70},
    lab/.style={font=\scriptsize},
    ticklab/.style={font=\scriptsize},
    leglab/.style={font=\scriptsize}
]

\def\W{5.6}
\def\H{3.0}
\def\S{2.8}
\def\C{1.6}

\begin{scope}[shift={(.3,0)}]
    \def\PVop{2.5}
    \def\NDop{1.3}

    \def\PVpe{0.25}
    \def\SUPpe{1.30} 

    \fill[chargegreen] (0,\NDop) rectangle (\S,\PVop);

    \fill[ndem] (0,\NDop) rectangle (\S,\C);

    \fill[dischargegreen] (\S,\PVpe) rectangle (\W,\SUPpe);

    \fill[ndem] (\S,\SUPpe) rectangle (\W,\C);

    \draw[axis] (0,0) -- (\W,0);
    \draw[axis] (0,0) -- (0,\H);
    \draw[split,dotted] (\S,0) -- (\S,\H);

    \draw[netline] (0,\NDop) -- (\S,\NDop);
    \draw[netline] (\S,\SUPpe) -- (\W,\SUPpe);

    \draw[cons] (0,\C) -- (\W,\C);
    \draw[pvline] (0,\PVop) -- (\S,\PVop);
    \draw[pvline] (\S,\PVpe) -- (\W,\PVpe);

    \node[lab] at ({\S/2},{(\NDop+\PVop)/2}) {Charge};
    \node[lab] at ({(\S+\W)/2},{(\PVpe+\SUPpe)/2}) {Discharge};

    \node[lab,rotate=90] at (-0.30,1.5) {kW};
    \node[ticklab] at ({\S/2},2.8) {\textbf{Off-Peak}};
    \node[ticklab] at ({(\S+\W)/2},2.8) {\textbf{Peak}};
    \node[ticklab] at ({\W/2},-0.28) {Time};
    \node[ticklab] at ({\W/2},3.2) {\textbf{Complementarity Case}};
    
\end{scope}

\begin{scope}[shift={(6.55,0)}]
    \def\NDop{.30}
    \def\PVop{1.4}
    \def\PVpe{1.15}
    \def\SUPpe{2.1125} 

    \fill[chargegreen] (0,\NDop) rectangle (\S,\PVop);

    \fill[ndem] (0,\C) rectangle (\S,\NDop);

    \fill[dischargegreen] (\S,\PVpe) rectangle (\W,\SUPpe);

    \fill[export] (\S,\SUPpe) rectangle (\W,\C);

    \draw[axis] (0,0) -- (\W,0);
    \draw[axis] (0,0) -- (0,\H);
    \draw[split,dotted] (\S,0) -- (\S,\H);

    \draw[netline] (0,\NDop) -- (\S,\NDop);
    \draw[netline] (\S,\SUPpe) -- (\W,\SUPpe);

    \draw[cons] (0,\C) -- (\W,\C);
    \draw[pvline] (0,\PVop) -- (\S,\PVop);
    \draw[pvline] (\S,\PVpe) -- (\W,\PVpe);

    \node[lab] at ({\S/2},{(\NDop+\PVop)/2}) {Charge};
    \node[lab] at ({(\S+\W)/2},{(\PVpe+\SUPpe)/2.45}) {Discharge};

    \node[lab,rotate=90] at (-0.30,1.5) {kW};
    \node[ticklab] at ({\S/2},2.8) {\textbf{Off-Peak}};
    \node[ticklab] at ({(\S+\W)/2},2.8) {\textbf{Peak}};
    \node[ticklab] at ({\W/2},-0.28) {Time};
    \node[ticklab] at ({\W/2},3.2) {\textbf{Substitution Case}};

    \begin{scope}[shift={(3.05,3.00)}]
        \fill[export] (0,-2.12) rectangle (0.40,-1.98);
        \draw[black] (0,-2.12) rectangle (0.40,-1.98);
        \node[anchor=west,leglab] at (0.32,-2.07) {Export};
        
        \fill[ndem] (0,-2.32) rectangle (0.40,-2.18);
        \draw[black] (0,-2.32) rectangle (0.40,-2.18);
        \node[anchor=west,leglab] at (0.32,-2.27) {Import};

        \draw[pvline] (0,-2.45) -- (0.40,-2.45);
        \node[anchor=west,leglab] at (0.32,-2.45) {PV Gen.};

        \draw[netline] (0,-2.65) -- (0.40,-2.65);
        \node[anchor=west,leglab] at (0.32,-2.65) {DER supply};

        \draw[cons] (0,-2.89) -- (0.40,-2.89);
        \node[anchor=west,leglab] at (0.32,-2.89) {Consumption};
    \end{scope}
\end{scope}

\end{tikzpicture}
\vspace{-0.2cm}
\caption{\small   \textbf{Left}: Storage increases the value of PV by switching the off-peak period to importing. \textbf{Right}: Storage decreases the value of PV by switching the peak period to exporting. Both plots depict a case $\pi_{\off}^-<\pi_{\off}^+<\eta^c\eta^d\pi_{\on}^-<\eta^c\eta^d\pi_{\on}^+$.}
    \label{fig:comp_sub}
\end{figure}
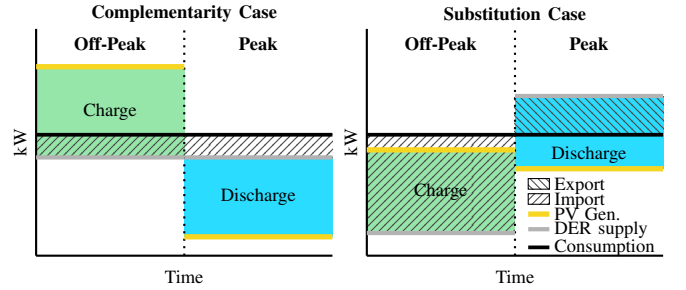
 Therefore, if only the off-peak boundary is crossed, PV and storage are complements; if only the peak boundary is crossed, they are substitutes. If both boundaries are crossed, they are complements when the weighted off-peak gain exceeds the weighted peak loss, substitutes when the reverse inequality holds, and locally neutral under equality. If neither boundary is crossed, storage does not change the marginal value of PV.

Before providing a complete characterization of complementarity, we introduce the following notation. Let
\[
\mathcal G_{\off}\!:=\![0,\bar g]\cap(g^{\off}\!(0),g^{\off}\!(z)),\ 
\mathcal G_{\on}\!:=\![0,\bar g]\cap(g^{\on}\!(z),g^{\on}\!(0)),
\]
denote the feasible PV capacities for which it is possible for storage to bring the corresponding period to net zero. Whether it does so optimally depends on the tariff. Using this notation, we provide the full conditions for whether PV and storage are complements or substitutes in the following theorem:

\begin{theorem}[PV--storage interaction]
\label{thm:pv_storage_interaction}
 For
$g\in[0,\bar g]\setminus\{g^{\off}(0),g^{\off}(z),g^{\on}(z),g^{\rm{nz}}\}$, let $\pi_b^{\alpha}=\pi_b^+$ if $g<g^{\rm{nz}}$ and $\pi_b^{\alpha}=\pi_b^-$ if $g>g^{\rm{nz}}$. Then
\begin{align}
&\Phi(g;z,\bm\pi)-\Phi(g;0,\bm\pi)\nonumber\\
&=\big[\min\{\pi_{\off}^+,\eta^c\eta^d\pi_{\on}^{\alpha}\}-\pi_{\off}^-\big]^+
\mathbbm{1}_{\{g \in \mathcal G_{\off}\}}\!\!\sum_{t\in\mathcal T_{\off}}\!\!\psi_t\nonumber\\
&\quad-\Big[\pi_{\on}^+-\max\Big\{\pi_{\on}^-,\frac{\pi_{\off}^{\alpha}}{\eta^c\eta^d}\Big\}\Big]^+
\mathbbm{1}_{\{g \in \mathcal G_{\on}\}}\!\!\sum_{t\in\mathcal T_{\on}}\!\!\psi_t.
\label{eqn:interaction}
\end{align}
PV and storage are complements, substitutes, or locally neutral according to the sign of the right-hand side of \eqref{eqn:interaction}. At the excluded capacities, the corresponding left- and right-hand comparisons are given by the one-sided limits of \eqref{eqn:interaction}. Proof is provided in \cite{lado&alahmed&botterud&amin:26JournalarXiv}
\end{theorem} 
\noindent
 We examine \eqref{eqn:interaction} by breaking it into three pairs of terms:
 
\noindent\textit{(1) PV thresholds:} The PV capacity required to be exactly at zero net demand in both periods, $g^{\rm{nz}}$, determines when the superscript $\alpha$ switches from $+$ to $-$. For $g<g^{\rm{nz}}$, the off-peak net zero boundary is met at lower charge levels, while for $g>g^{\rm{nz}}$, the peak boundary is met at lower charge levels. Whether storage reaches either boundary depends on its capacity $z$, through the indicator variables, and the associated rates.

Condition $g\in\mathcal{G}_{\off}$ indicates that PV produces off-peak exports and that storage capacity is sufficient to eliminate those exports (illustrated in the left-hand panel of Fig. \ref{fig:comp_sub}). Condition $g\in\mathcal{G}_{\on}$ indicates that peak imports occur without storage and that storage capacity is sufficient to eliminate those imports (illustrated in the right-hand panel of Fig. \ref{fig:comp_sub}).


\noindent\textit{(3) PV generation weights:} The factors $\Psi_{\off}$ and $\Psi_{\on}$ weight the off-peak gain and peak loss by the generation from an additional unit of PV capacity in each period. Holding the price terms and indicators fixed, a smaller $\Psi_{\on}/\Psi_{\off}$ gives less weight to substitution relative to complementarity.

\noindent\textit{(3) Price spreads:} The brackets capture the change in marginal value when either period switches between importing and exporting. Thus, price spreads and the battery efficiency shape storage's increases PV's marginal value in the off-peak more than it decreases marginal value in the peak.
The role of each price spread is clearest when examined individually. 
\begin{corollary}[Import--export price spread]
\label{cor:pv_storage_price_gaps}
With only one price spread either the peak or off-peak component is zero, and the other has a known sign:
\begin{align*}
\pi_{\off}^+=\pi_{\off}^-
,\ 
\pi_{\on}^+=\pi_{\on}^-
&\Longrightarrow
\Phi(g;z,\bm\pi)=\Phi(g;0,\bm\pi),\\
\pi_{\off}^+>\pi_{\off}^-
,\ 
\pi_{\on}^+=\pi_{\on}^-
&\Longrightarrow
\Phi(g;z,\bm\pi)\geq\Phi(g;0,\bm\pi),\\
\pi_{\off}^+=\pi_{\off}^- ,\ \pi_{\on}^+>\pi_{\on}^-
&\Longrightarrow
\Phi(g;z,\bm\pi)\leq\Phi(g;0,\bm\pi).
\end{align*}
The complementarity relation is strict if and only if
$g\in G_{\off}
\text{ and }
\eta^c\eta^d\pi_{\on}^+>\pi_{\off}^-,
$
and the substitution relation is strict if and only if $g\in G_{\on}
\quad\text{and}\quad
\eta^c\eta^d\pi_{\on}^+>\pi_{\off}^+.
$
At PV capacities where the marginal value is not unique, the same conclusions apply separately to the corresponding left-and right-hand marginal values, with strictness determined by the adjacent open interval. Proof is provided in \cite{lado&alahmed&botterud&amin:26JournalarXiv}.
\end{corollary} 
Thus, price gap only off-peak produces complementarity, a price gap only peak produces substitution, and no price gap in either period produces no interaction between PV and storage.

Finally, we connect PV marginal value differences to optimal PV capacities. We denote a selected privately optimal investment pair for $\bar g>0$ by $(g^{\rm DER}(\bm\pi,\bar z),z^{\rm DER}(\bm\pi,\bar z))$, suppressing its dependence on $\bar g$. If multiple pairs are optimal, we select one with the largest PV capacity and any associated optimal storage capacity may be chosen. We now show the relationship between marginal values with and without storage and PV capacities :
\begin{subequations}\begin{proposition}[Marginal PV value and optimal investment]
\label{prop:marginal_value_investment}
If $g^{\rm DER}(\bm\pi,0)>0$ and $g^{\rm DER}(\bm\pi,\bar z)<\bar g$, either of the inequalities
\begin{align*}
\overline{\Phi}\left(g^{\rm DER}(\bm\pi,0);z^{\rm DER}(\bm\pi,\bar z),\bm\pi\right)&\geq\overline{\Phi}\left(g^{\rm DER}(\bm\pi,0);0,\bm\pi\right),\\
\underline{\Phi}\left(g^{\rm DER}(\bm\pi,\bar z);z^{\rm DER}(\bm\pi,\bar z),\bm\pi\right)&\geq\underline{\Phi}\left(g^{\rm DER}(\bm\pi,\bar z);0,\bm\pi\right)
\end{align*}
implies $g^{\rm DER}(\bm\pi,\bar z)\geq g^{\rm DER}(\bm\pi,0)$.
\end{proposition}\end{subequations}
\noindent Prop.~\ref{prop:marginal_value_investment} connects the marginal-value comparisons above to investment: storage need not increase PV value at every capacity to increase optimal PV investment. It is sufficient that PV marginal value is weakly greater at either the with-storage or no-storage optimum.
Thus, prices determine how much charge and discharge can profitably occur. PV capacity and storage capacity, along with the demand, cost, and capacity factor parameters, determine which boundaries are crossed.

\noindent\textbf{TI and Symmetric Pricing - Practically Relevant Special Cases:} Under TI pricing, $\pi_{\on}^+=\pi_{\off}^+$ and $\pi_{\on}^-=\pi_{\off}^-$, the tariff lies in investment regions A or B. Storage is either not profitable (Region A) or built to avoid exporting in either period (Region B). FA symmetric ToU tariff, $\pi_{\on}^+=\pi_{\on}^-$ and $\pi_{\off}^+=\pi_{\off}^-$, lies within investment Region A if the round trip efficiency is less than the off-peak to peak price ration and investment Region F otherwise. Because the tariff is symmetric, the marginal values do not change when switching between importing and exporting, so  PV and storage do not affect each others value (Cor. \ref{cor:pv_storage_price_gaps}). each investment is zero or at its maximum unless its marginal value equals its investment cost. At equality, all capacities are optimal. Finally, a legacy TI-symmetric tariff lies in Region A, induces either zero storage investment and zero or full PV investment, except at break-even cost.

\section{DU's Net Cost of Service} \label{sec:system}
In this section, we examine how PV and storage adoption affect the DU's net cost of serving a prosumer.
We denote the DU's period dependent costs, including  energy and non-energy costs attributable to netting period $b$, by $c_b\in\mathbbm{R}_+$ for $b\in\{\on,\off\}$ (more details on attributable costs are provided in Section \ref{sec:CaseStudy}). Let $C^{\rm{fix}}\! \in \mathbb{R}_+$ denote the fixed cost associated with serving a customer and $\bm c$ denote the vector composed of attributable and fixed costs.
The DU's {\em net cost of serving a customer} is given by
\begin{align*}
C(\bm d^+,\bm d^-;\bm\pi):=
\sum_{b\in\mathcal B}c_b(d_b^+-d_b^-) +C^{\rm fix} -P^{\mbox{\tiny NEM}}(\bm d^+,\bm d^-;\bm\pi).
\end{align*}
where the first two terms are the DU's varying and fixed costs, and the third term is its revenue. For given investments $(g,z)$, we write
$C(g,z;\bm\pi) :=C\bigl(\bm d^{+*}(g,z;\bm\pi), \bm d^{-*}(g,z;\bm\pi);\bm\pi\bigr),$
where the starred quantities denote a selected optimal operating
response conditional on $(g,z)$.Thus, the DU fully recovers its cost of serving the prosumer when $C(g,z;\bm\pi)\!=\! 0$. Moreover, as net demand decisions are affected by DER investments, a prosumer's bill savings depend on their PV and storage decisions. It is useful to decompose that effect. For every admissible tariff and every feasible $(g,z)$,
\begin{align}
&C(g,z;\bm\pi)-C(0,0;\bm\pi)\label{eqn:net_cost_decomp}\\
&=\![\Pi(0,\!0;\bm\pi)\!-\!\Pi(g,\! z;\! \bm\pi)]
\nonumber+\!g(c^g\!-\!\sum_{b\in\mathcal B}c_b\sum_{t\in\mathcal T_b}\psi_t)
\\\nonumber&\hspace{9pt}+\!(c^zz\!-\!T(\eta^dc_{\on}-c_{\off}/\eta^c)\frac{\eta^cQ^{c*}_{\off}}{T}),
\nonumber
\end{align}
where $Q^{c*}_{\off}$ is total grid-side off-peak charge under optimal operation conditional on $(g,z)$. The first term is the prosumer's private gain from adoption, which is nonnegative at jointly optimal investments because $(0,0)$ is feasible for the prosumer; the second and third are the gaps between investment cost and attributable value. At jointly optimal investments $(g^*,z^*)$, positive storage cost implies full utilization, so $\eta^c Q^{c*}_{\off}/T=z^*$. The last two terms therefore become
\[g^*(c^g-\sum_{b\in\mathcal B}c_b\Psi_b)
+z^*(c^z-T(\eta^dc_{\on}-c_{\off}/\eta^c)).\]

 Using the notation introduced in Sec. \ref{sec:invest}, let
$C^{\rm DER}(\bm\pi,\bar z) := C(g^{\rm DER}(\bm\pi,\bar z), z^{\rm DER}(\bm\pi,\bar z);\bm\pi).$
We first compare the effects of DER adoption on net cost relative to the no-DER case, $C^{\rm 0}(\bm\pi,0)=C(0,0;\bm\pi)$:
\begin{proposition}[DER Adoption and DU Net Costs]
\label{prop:prescribed_pv_cost_recovery}
Suppose $\bar z=0$ and
\begin{align}
c^g>\sum_{b\in\mathcal B}c_b
\Psi_b.
\label{eq:fixed_tariff_pv_cost_condition}
\end{align}
then $C^{\rm{DER}}(\bm\pi,0)\geq C^{\mathrm{0}}(\bm\pi,0)$. If storage investment is allowed, $\bar{z} >0$ for any $i\in\mathcal{N}$, also suppose
\begin{align}
c^z&>T\Big(\eta^dc_{\on}
-\frac{c_{\off}}{\eta^c}\Big),
\label{eq:fixed_tariff_storage_cost_condition}
\end{align}
Then, 
$C^{\rm{DER}}(\bm\pi,\bar{z})\geq C^{\mathrm{0}}(\bm\pi,0)$ for every $\bm\pi$.
For each tariff, the corresponding inequality is strict whenever optimal PV investment is positive.
Proof is provided in \cite{lado&alahmed&botterud&amin:26JournalarXiv}.
\end{proposition}

\noindent When PV costs more than the capacity-factor-weighted attributable costs, any tariff that induces adoption raises the DU's net cost of serving the prosumer.
When both conditions hold, PV and storage investment weakly increase DU net cost relative to the no-DER setting, with a strict increase whenever there is positive investment in either technology.

 Prop. \ref{prop:prescribed_pv_cost_recovery} does not compare the PV alone against the joint PV and storage case. This comparison depends on the interaction between PV and storage, which we discuss next:
\begin{proposition}[Complementarity and DU net cost at a fixed tariff]
\label{prop:fixed_tariff_storage_net_cost}
Let $\widehat{\bm\pi}:=(\pi^+_{\off},\pi^-_{\off},\widehat\pi^+_{\on},\widehat\pi^-_{\on})$ represent the corresponding storage-cost-adjusted tariff and suppose optimal PV capacities with and without storage are unique. If either of the following inequalities holds:\begin{subequations}
\begin{align*}
\overline{\Phi}\!\left(g^{\rm DER}(\bm\pi,0);z^{\rm DER}(\bm\pi,\bar z),\bm\pi\right)
&\geq
\overline{\Phi}\!\left(g^{\rm DER}(\bm\pi,0);0,\bm\pi\right),
\\
\underline{\Phi}\!\left(g^{\rm DER}(\bm\pi,\bar z);z^{\rm DER}(\bm\pi,\bar z),\bm\pi\right)
&\geq
\underline{\Phi}\!\left(g^{\rm DER}(\bm\pi,\bar z);0,\bm\pi\right),
\end{align*}\end{subequations}
in addition to \eqref{eq:fixed_tariff_pv_cost_condition} and \eqref{eq:fixed_tariff_storage_cost_condition}. Then $g^{\rm{DER}}(\bm\pi,\bar z)\!\geq\! g^{\rm{DER}}(\bm\pi,0)$ and $C^{\rm{DER}}(\bm\pi,\bar z)\geq C^{\rm{DER}}(\bm\pi,0)$. This inequality is strict when storage investment is positive. Proof is provided in \cite{lado&alahmed&botterud&amin:26JournalarXiv}.
\end{proposition}

\noindent  Thus, under the complementarity and cost conditions of Prop. \ref{prop:fixed_tariff_storage_net_cost}, the introduction of storage increases both optimal PV and the net cost associated with a serving a prosumer. 

This result highlights a tradeoff for regulators. At a fixed tariff $\bm\pi$ and under cost conditions \eqref{eq:fixed_tariff_pv_cost_condition}, \eqref{eq:fixed_tariff_storage_cost_condition}, complementarity between PV and storage increases the DU's net cost of serving adopters and, under class-wide cost recovery, increases the costs recovered from other customers. At the same time, complementarity raises the private value of PV. In particular, when
$g^{\rm{DER}}(\bar z,\bm\pi)\geq g^{\rm{DER}}(0,\bm\pi)$,
storage can preserve PV capacity under tariffs that would otherwise reduce PV investment. Thus, the same interaction that worsens cost recovery can also make PV investment compatible with tariffs that provide weaker incentives for standalone PV. This tradeoff is particularly relevant when maintaining distributed PV deployment is valuable because utility-scale alternatives face constraints such as interconnection delays or limited available land.

In Sect. \ref{sec:CaseStudy}, we examine the effect of DERs on net cost in a multi-netting period setting using data from Massachusetts. We decompose the effects roles of complementarity, substitution and their interaction with technology costs.

\section{Case Study - Massachusetts Prosumers}\label{sec:CaseStudy} 
We conduct a case study using a population of 500 Massachusetts single-family homes under NEM. In this case study, we include capacity factor uncertainty and a greater number of netting periods. We illustrate the mechanisms underlying our theoretical findings in this broader setting.
\subsection{Parameter Derivations:}
\noindent\textbf{\textit{Customer and Technology Cost Parameters:}} To generate the experiments in our case study, we use simulated hourly demand profiles and PV generation data for 2018 from the ResStock dataset, aggregating hourly demands over each netting period \cite{osti_2583488}. 

Solar and storage costs use baseline gross installed-cost assumptions of 2{,}950 \$/\text{kW}$ for PV and 972 \$/\text{kWh}$ for battery storage, expressed in 2025 dollars and informed by interpolating NLR benchmark cost data from 2023 and projections to 2035 to 2028 \cite{nrel:2024atb}. These costs are annualized using monthly payments over a 10-year loan at 5.5 percent APR, without tax credits. This results in $c^g=384.18$ \$/kW-year and $c^z=126.59$ \$/kWh-year. 

\noindent\textbf{\textit{PV Uncertainty:}} We model PV uncertainty by multiplying each household's 2018 ResStock PV profile by a mean-one, scaled beta-distributed weather factor. For each calendar month and peak or off-peak period, we calibrate the factor's standard deviation relative to its mean using year-to-year variation in ISO-NE's normalized PV production across the three Massachusetts zones over 2014--2025.\footnote{ISO New England, \href{https://www.iso-ne.com/static-assets/documents/2020/07/btm_pv_data.xlsx}{normalized PV data} and \href{https://www.iso-ne.com/static-assets/documents/2020/07/btm_pv_data_documentation.pdf}{documentation}.} We represent each period's distribution with 10 unequally weighted scenarios. In each scenario, households share the same weather-factor realization while retaining their distinct ResStock profiles. Investment and charge/discharge decisions are made before uncertainty is realized. Further details can be found in \cite{lado&alahmed&botterud&amin:26JournalarXiv}.

\noindent\textbf{\textit{Cost Estimates:}}
We calculate peak and off-peak costs for each month. Each peak or off-peak netting period has an attributable cost rate $c_b$, comprising energy~(LMP), capacity costs from the Forward Capacity Market~(FCM), Regional Network Service~(RNS) and Local Network Service~(LNS) costs, losses, environmental compliance, and distribution. Thus,
\begin{align*}
c_b=c_b^{\mathrm{LMP}}+c_b^{\mathrm{FCM}}+c_b^{\mathrm{RNS}}+c_b^{\mathrm{LNS}}+c_b^{\mathrm{loss}}+c_b^{\mathrm{comp}}+c_b^{\mathrm{dist}}.
\end{align*}
We retain 2018 load profiles and timing while using 2021--2025 cost levels expressed in 2025 dollars. We use the 2021--2025 cost levels as 2018 FCM costs are an unrepresentative outlier. The details regarding the attributable costs calculations are in \cite{lado&alahmed&botterud&amin:26JournalarXiv}. 
Our seasons are summer (June--Sept.), winter (Oct.--May.). The peak periods we use are 4--8 p.m. for summer and 5--9 p.m. for non-summer months.
Across the year, PV avoids attributable costs of $17.95$ \$/kW-year during peak periods and $124.12$ \$/kW-year during off-peak periods. In total, PV only avoids around one third of its investment cost.
Storage's average annual avoided attributable cost of \$79.19/kWh-year falls short of its annualized investment cost by \$47.39/kWh-year. The PV and storage cost comparisons underlying \eqref{eq:fixed_tariff_pv_cost_condition} and \eqref{eq:fixed_tariff_storage_cost_condition} both hold. Positive PV-only adoption therefore increases DU net cost relative to no adoption; the effect of joint adoption also depends on the interaction between PV and storage.

We next examine how PV and storage interaction affects DU net cost and draw lessons for tariff design. We examine three tariffs: Tariff P has an import--export spread only during peak hours, Tariff O only during off-peak hours, and Tariff B includes both price spreads. In the order $(\pi_{\on}^{+},\pi_{\on}^{-},\pi_{\off}^{+},\pi_{\off}^{-})$, their rates at the end of the paths are $(0.760,0.440,0.310,0.310)$, $(0.760,0.760,0.310,0.190)$, and $(0.760,0.440,0.310,0.190)$ respectively (\$/kWh). The import rates are based upon a symmetric TOU tariff loosely based on a proposal by the Massachusetts Department of Energy Resources.\footnote{Massachusetts Department of Energy Resources, \href{https://www.mass.gov/doc/rate-task-force-ratemaking-straw-proposal/download}{Ratemaking Straw Proposal}.} We add price asymmetries and altered the peak and off-peak rates to highlight how asymmetries in each period cause complementarity and substitution.

\noindent\textbf{\textit{Effect of PV Capacity on DER Interaction:}}
In Thm. \ref{thm:pv_storage_interaction}, the combination between DER capacities and price spread form determine the presence of these effects while relative generation of each period weights each effect. To highlight these factors, we  compare the marginal values of PV at different PV capacity levels under Tariff B during July. We select one prosumer, vary their PV capacity between 0 to 15 kWs, fix their storage capacity to $10$ kWh. Tariff B is asymmetric in both the peak and off-peak periods, implying it can potentially induce complementarity or substitution.  

The left-hand panel of Fig. \ref{fig:fixedinvest} displays the results of this experiment in the deterministic setting with capacity factors at their mean values. The substitutive effect begins as capacity crosses $g^{\on}(z)$, the infimum of $\mathcal{G}^{\on}$. Substitution here is possible only because of the price spread between peak import and export rates. As capacity increases, it enters $\mathcal{G}^{\off}$, which lies between $g^{\off}(0)$ and $g^{\off}(z)$, and complementarity is observed. Substitution still exists, but because the off-peak has a more generation than the peak, the complementarity effect dominates. Note that the price spreads are equal here so the difference comes from the relative generation. Leaving this region, substitution is again the only effect. Threshold $g^{\off}(0)$ lies above $\bar{g}=15$ and is not plotted.
\begin{figure}
    \centering
    \includegraphics[width=1\linewidth]{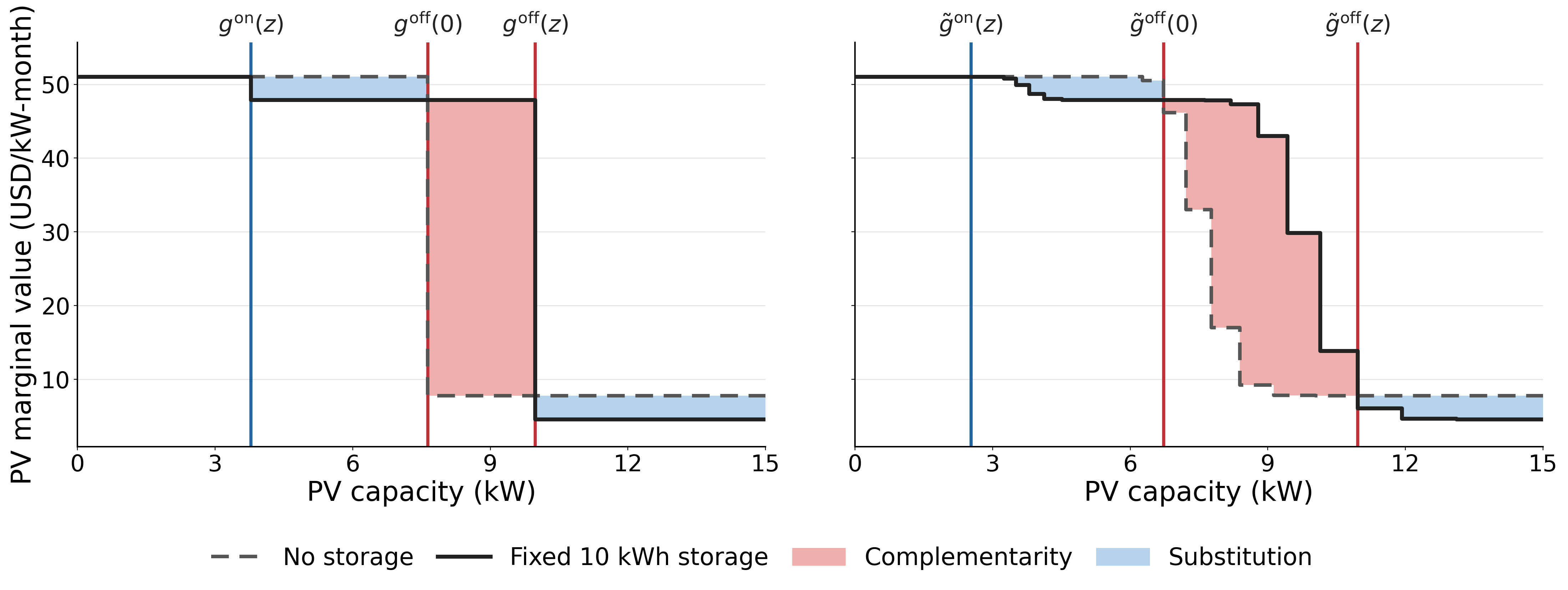}
    \caption{\small Interaction effects as a function of PV capacity. \textbf{Left:} Marginal values with deterministic capacity factors. \textbf{Right:} Marginal values with stochastic capacity factors.}
    \label{fig:fixedinvest}
\end{figure}
The right-hand panel of Fig. \ref{fig:fixedinvest} shows similar effects, but in the stochastic setting. Here, $\tilde g^{\on}(z)$, $\tilde g^{\off}(0)$, and $\tilde g^{\off}(z)$ replace the deterministic threshold. They retain the numerators from \eqref{eqn:glevel_1}, but $\Psi_{\on}$ or $\Psi_{\off}$ is replaced, respectively, with
\begin{align*}
&\inf\left\{\psi:\mathbb{E}\!\left[\Psi_{\off}\mathbf{1}\{\Psi_{\off}\le\psi\}\right]\ge(1-q)\mathbb{E}[\Psi_{\off}]\right\}
\\
&\inf\left\{\psi:\mathbb{E}\!\left[\Psi_{\off}\mathbf{1}\{\Psi_{\off}\le\psi\}\right]\geq q\mathbb{E}[\Psi_{\off}]\right\},\\
&\inf\left\{\psi:= \mathbb E\left[\Psi_{\on}\mathbf 1\{\Psi_{\on}\leq\psi\}\right]\geq \mathbb E[\Psi_{\on}]
\right\}
\end{align*}
where $q=\frac{(\pi_{\on}^{+}-\pi_{\on}^{-})\mathbb{E}[\Psi_{\on}]}{(\pi_{\off}^{+}-\pi_{\off}^{-})\mathbb{E}[\Psi_{\off}]}$. Under uncertainty, different production scenarios reach net zero at different PV capacities. The thresholds use production-weighted quantiles because scenarios in which an additional kW generates more electricity have a larger effect on expected marginal PV value. While the exact form of the transitions between complementarity and substitution change, the reasons why each appears in each region remains the same.


\noindent\textbf{\textit{Import--Export Price Spreads:}} We now illustrate the effects of the spread in each period, shown in Thm.~\ref{thm:pv_storage_interaction} and Cor.~\ref{cor:pv_storage_price_gaps}. We plot investment decisions, changes to imports and exports and DU net as we shift from the common symmetric tariff to Tariffs O or P. To reach Tariff O, we begin with $\pi_{\off}^{+}=\pi_{\off}^{-}=0.310$ \$/kWh and lower the off-peak export price to $\pi_{\off}^{-}=0.310-\delta$ \$/kWh. To reach Tariff P, we begin with $\pi_{\on}^{+}=\pi_{\on}^{-}=0.760$ \$/kWh and lower the peak export price to $\pi_{\on}^{-}=0.760-\delta$ \$/kWh. We evaluate 121 equally spaced values of $\delta\in[0,0.12]$ for Tariff O and 161 equally spaced values of $\delta\in[0,0.31]$ for Tariff P. As we change each export rate, all other rates remain fixed. At each value, we optimize the investment and operations of all prosumers.
\begin{figure}
    \centering
    \includegraphics[width=1\linewidth]{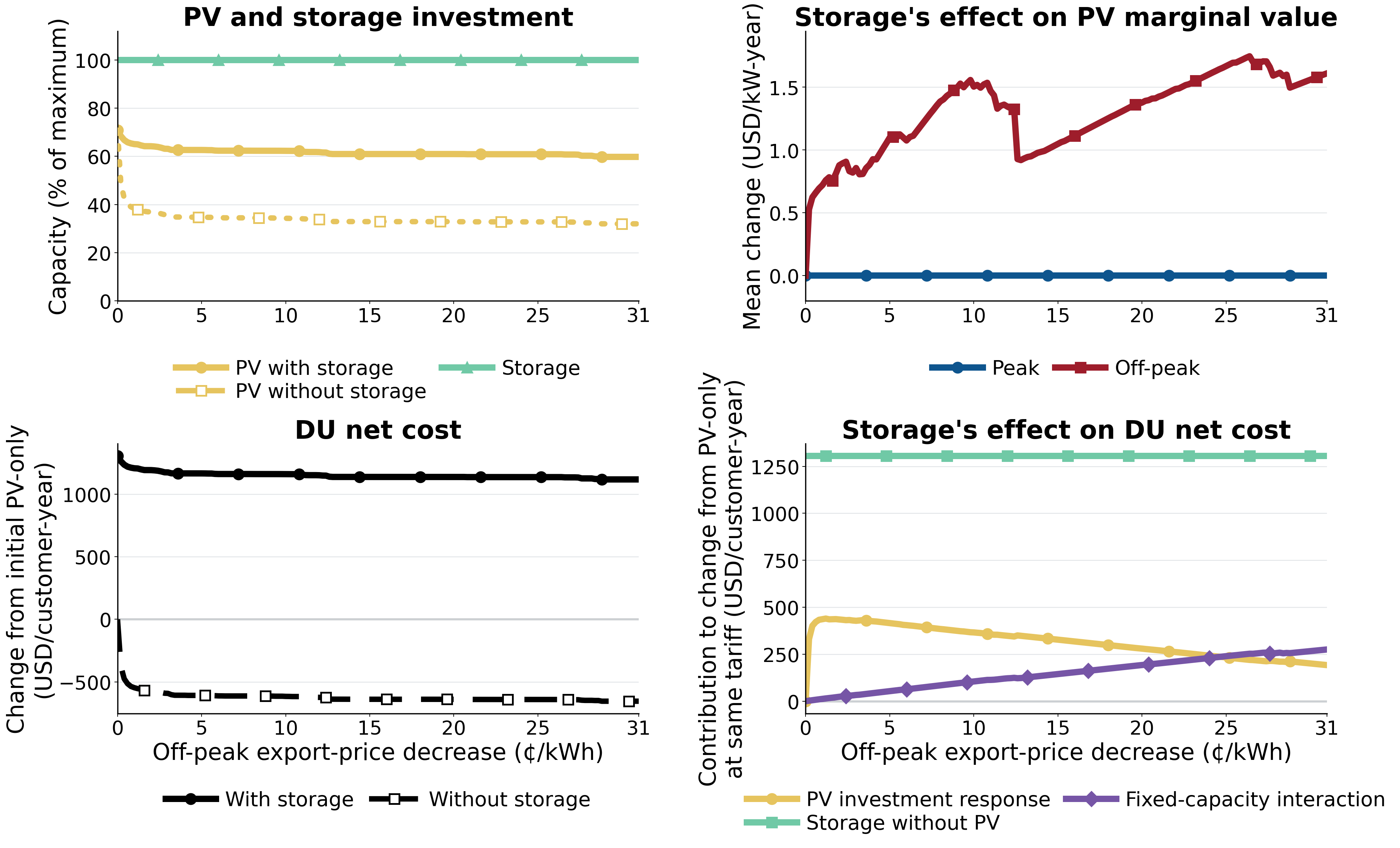}
    \caption{\small Effects of a decrease in the off-peak export price. \textbf{Top Left:} Optimal PV and storage investments, expressed as percentages of their respective maximum capacities. \textbf{Top Right:} The effect of storage on PV marginal value by period.  \textbf{Bottom Left:} DU net-cost changes relative to PV alone at $\delta=0$. \textbf{Bottom Right:} Effect of PV, storage, and their interaction on DU net costs.}
    \label{fig:complementarity}
\end{figure}

Fig.~\ref{fig:complementarity} shows how widening the off-peak import--export spread affects optimal investment, import and export, and the effects on the DU's net cost. Complementarity is evident in the top left panel by the difference in PV capacity with storage (solid yellow) and without storage (dotted yellow). Storage (green) is at maximum at all points. As off-peak export compensation falls, PV investment without storage declines from 70.9\% to 32.0\%. With storage, PV only declines to 59.8\%. The top right panel corroborates this interpretation, showing that the effect of storage on the PV marginal value is only in the off-peak, where the price perturbation occurs. In the bottom left panel, lowering the off-peak export rate decreases the net cost of DU, but this decrease is much greater when PV is not paired with storage. This difference, between the the DU cost with storage (solid black) and (dashed black), is decomposed in the bottom right panel. By increasing PV capacity (yellow), its own direct effect on costs (green), and by raising PV marginal value at each endogenous capacity (purple). Complementarity sustains PV investment but at the expense of the improvement in DU cost recovery.
\begin{figure}
    \centering
    \includegraphics[width=1\linewidth]{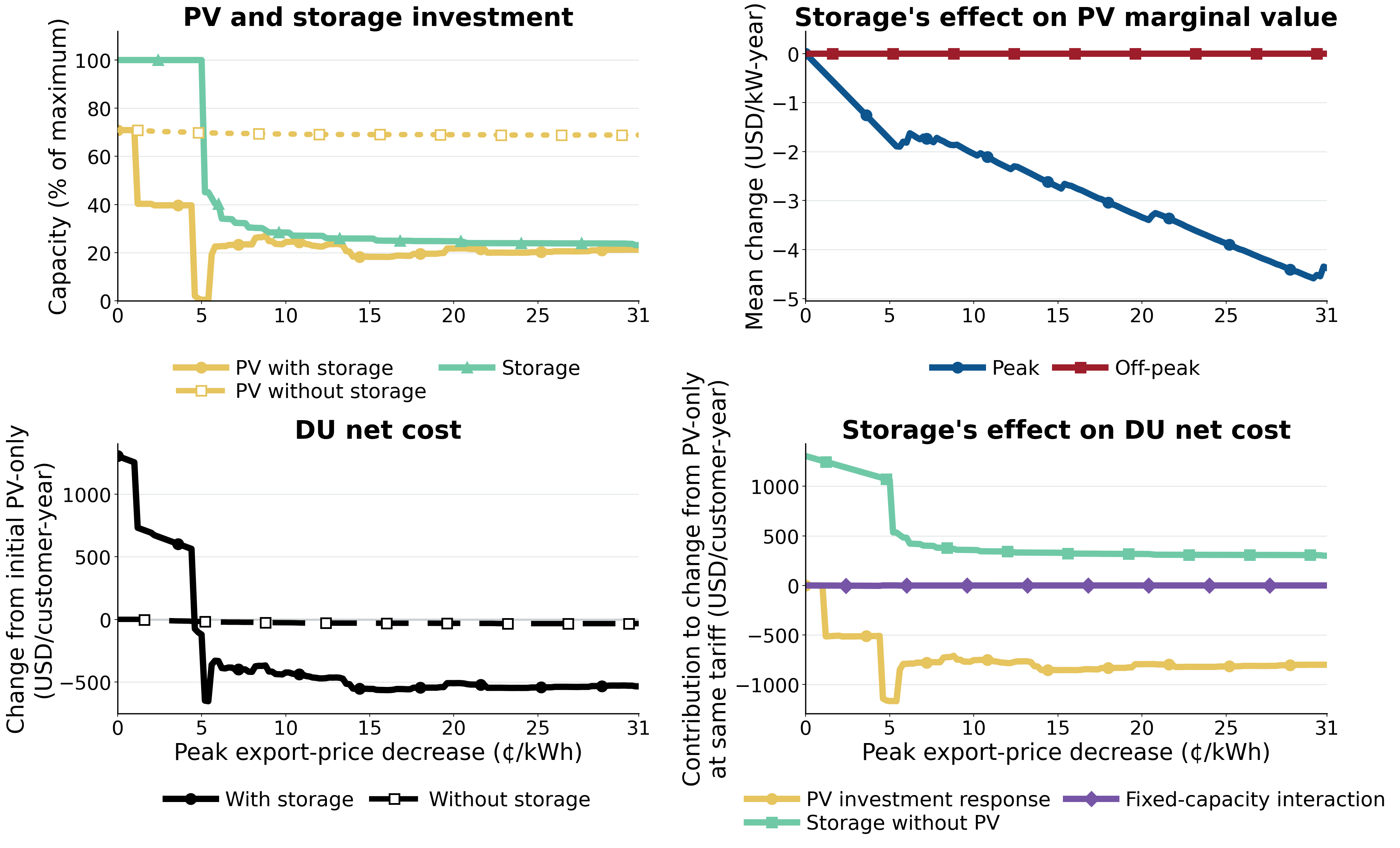}
    \caption{\small Decrease in the peak export price. \textbf{Top Left:} Optimal PV and storage investments, expressed as percentages of their respective maximum capacities. \textbf{Top Right:} The effect of storage on PV marginal value by period.  \textbf{Bottom Left:} DU net-cost changes relative to PV alone at $\delta=0$. \textbf{Bottom Right:} Effect of PV, storage, and their interaction on DU net costs.}
    \label{fig:substitution}
\end{figure}

Fig.~\ref{fig:substitution} shows the analogous effect of reducing peak export compensation. In the top left, substitution is evident in the difference between PV capacities with and without storage. PV investment without storage changes little, dropping from 70.9\% to 68.7\% of maximum capacity. With storage, PV falls to 26.5\% and storage falls from 100\% to 22.7\%. The top-right panel links this difference to storage's reduction in peak-period PV marginal value. As storage investment falls, this reduction can weaken and PV investment can rebound, producing the nonmonotonic response of PV capacity in the top-left panel.
In the bottom left, DU net cost falls by about $\$981$ per customer-year with PV and storage, compared with only $\$36$ for PV without storage. The bottom-right panel shows that storage directly increases DU net cost, but reduced PV investment more than offsets this increase at smaller export prices. Although the fixed-capacity interaction term is small at the chosen capacities, substitution substantially affects DU net cost through the PV investment response. PV rebounds cause local cost increases when their cost contribution exceeds the savings from reduced storage. Thus, substitution can be associated with both higher and lower DU net costs depending on the price asymmetry.

\noindent\textbf{\textit{PV and Storage Investment Costs:}} 
We next examine how DER investment costs affect DER interaction and DU net cost. Gray vertical dashed lines mark the baseline costs. The top row of Fig.~\ref{fig:PV_sensitivity} varies PV cost under Tariff O, which induces complementarity, and Tariff P, which induces substitution. PV adoption begins near the same cost under both tariffs. Under Tariff O, storage is always at maximum capacity. PV with storage is well above PV alone for PV costs below $380$ \$/kW. PV both with and without storage rises sharply to maximum capacity as its cost falls from about $\$384$ to $\$352/\mathrm{kW\text{-}year}$, while storage declines slightly. The high storage capacity under Tariff O results in higher DU net costs for PV and storage combined with PV alone and both PV alone and PV with storage under Tariff P. In contrast, there is lower DU net cost under Tariff P than Tariff O for low PV costs when there is no storage. Finally, the effect of storage on DU costs is much smaller under Tariff P than Tariff O, as more PV causes a decrease in storage capacity and related cost shifts.
\begin{figure}
    \centering
    \includegraphics[width=1\linewidth]{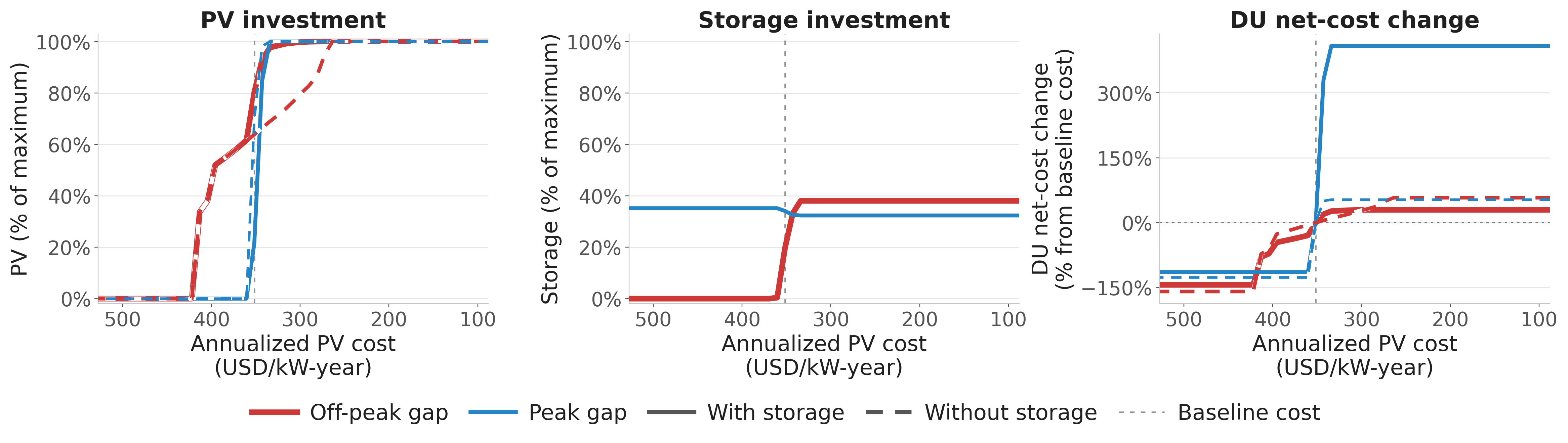}
    \caption{\small The top and bottom rows vary annualized PV and storage costs, respectively, holding the other cost fixed. \textbf{Left:} Optimal PV investment with and without storage for Tariff-O and Tariff-P. \textbf{Center:} Optimal joint storage investment for both tariffs. \textbf{Right:} DU net-cost change relative to the no-DER baseline, with and without storage.}
    \label{fig:PV_sensitivity}
\end{figure}

The bottom row varies storage cost while holding PV cost fixed. Under Tariff O, reducing storage cost from $\$148$ to $\$143/\mathrm{kWh\text{-}year}$ raises joint PV from 32.0\% to 59.8\% (left) and storage investment from zero to maximum capacity (center). On the right, DU net cost rises by about $\$1{,}772$ per customer-year relative to PV alone. In contrast, storage adoption displaces PV under Tariff P. Joint PV falls from 68.9\% to zero as storage cost declines from $\$143/\mathrm{kWh\text{-}year}$ as storage increases from zero to maximum capacity. DU net cost falls as PV is displaced by storage. Cheaper storage supports PV under Tariff O but displaces it under Tariff P, with opposite consequences for DU cost recovery. As the tariffs are fixed, the exact effects are very tariff dependent.



\section{Conclusion}
This paper characterizes the optimal joint investment and operational decisions of a prosumer who can invest in BTM PV and storage under a NEM tariff with peak and off-peak prices. From these optimal decisions, we isolate the factors that cause PV and storage to act as complements or substitutes and examine the implications for DU cost recovery. 

PV and storage are complements when storage increases PV value in the off-peak more than it decreases PV marginal value in the peak. The three factors that contribute to this are the ratio of capacity factors in the peak and off-peak, the export-import price spread in each period, and DER capacities. Higher off-peak capacity factors and price spreads contribute to complementarity, while those same factors contribute to substitution during the peak. Each effect can only exist when there are PV and storage capacities capable of reaching the net net-zero exchange threshold in either period. This understanding of DER interaction has implications for the cost recovery and decarbonization dilemma faced by regulators as DERs become cheaper and more widespread. Under a complementarity inducing tariff, storage can increase the cost shifts caused by PV. At the same time, complementarity can increase the penetration of PV and contribute to the decarbonization of the electricity grid.

These findings suggest that regulators should evaluate PV and storage jointly. Policies that promote co-adoption can support PV deployment while increasing DU revenue shortfalls. However, tariffs that would previously not be considered do to their negative effect on PV could provide an appropriate balance between balancing costs and PV adoption. Conversely, storage that displaces PV can improve cost recovery while reducing PV deployment. These outcomes depend on the timing of import-export price asymmetries and prevailing technology costs. Tariff design should therefore weigh storage's effect on PV investment alongside its operational benefits, balancing decarbonization objectives with utility cost recovery.
\label{sec:conclusion}

\bibliographystyle{IEEEtranDOI}
\bibliography{references}

\end{document}